\documentclass[aps,pra]{revtex4}

\usepackage{amsmath}
\usepackage{graphicx}
\usepackage{dcolumn}
\usepackage{longtable}

\newcommand{\be}{\begin{equation}}
\newcommand{\ee}{\end{equation}}
\newcommand{\bearray}{\begin{eqnarray}}
\newcommand{\eearray}{\end{eqnarray}}
\newcommand{\bse}{\begin{subequations}}
\newcommand{\ese}{\end{subequations}}

\begin{document}

\title{Three-dimensional Fourier transforms, integrals of spherical Bessel functions, and novel delta function identities}

\author{Gregory S. Adkins}
\email[]{gadkins@fandm.edu}
\affiliation{Franklin \& Marshall College, Lancaster, Pennsylvania 17604}
\date{\today}

\begin{abstract}
We present a general approach for evaluating a large variety of three-dimensional Fourier transforms.  The transforms considered include the useful cases of the Coulomb and dipole potentials, and include situations where the transforms are singular and involve terms proportional to the Dirac delta function $\delta(\vec r\,)$.  Our approach makes use of the Rayleigh expansion of $\exp({i \vec p \cdot \vec r\, })$ in terms of spherical Bessel functions, and we study a number of integrals, including singular integrals, involving a power of the independent variable times a spherical Bessel function.  We work through several examples of three-dimensional Fourier transforms using our approach and show how to derive a number of identities involving multiple derivatives of $1/r$, $1/r^2$, and $\delta(\vec r\,)$.
\end{abstract}

\maketitle

%%%%%%%%%%%%%%%%%%%%%%%%%%%%%%%%%%%%%%%%%%%%%%%%%%%

\section{Introduction}
\label{introduction}

Performing three-dimensional Fourier transforms is a routine task in physics research and in the upper-level physics classroom.  A common problem is to transform two-body interaction operators from momentum space to coordinate space and back.  Such operators might originate in a momentum-space Feynman diagram analysis but be most conveniently evaluated in coordinate space.  As useful and representative examples, consider the following transforms from the text of Berestetskii, Lifshitz, and Pitaevskii \cite{Berestetskii82} in their derivation of the Breit equations \cite{Breit29} for electron-electron and electron-positron interactions:
\bse \label{FT}
\bearray
I_a &=& \int \frac{d^3 p}{(2 \pi)^3} e^{i \vec p \cdot \vec r} \, \frac{1}{p^2} = \frac{1}{4 \pi r} , \label{FTa} \\
I_b &=& \int \frac{d^3 p}{(2 \pi)^3} e^{i \vec p \cdot \vec r} \, \frac{p_i}{p^2} = \frac{i x_i}{4 \pi r^3} , \label{FTb} \\
I_c &=& \int \frac{d^3 p}{(2 \pi)^3} e^{i \vec p \cdot \vec r} \, \frac{p_i p_j}{p^4} = \frac{1}{8 \pi r} \left ( \delta_{i j}-\hat x_i \hat x_j \right ) , \label{FTc} \\
I_d &=& \int \frac{d^3 p}{(2 \pi)^3} e^{i \vec p \cdot \vec r} \, \frac{p_i p_j}{p^2} = \frac{1}{3} \delta_{i j} \delta(\vec r \,) + \frac{1}{4 \pi r^3} \left ( \delta_{i j} - 3 \hat x_i \hat x_j \right ) . \label{FTd} 
\eearray
\ese
The position vector has components  $\vec r = (x_1,x_2,x_3)=(x, y, z)$, and $\hat x_i = x_i/r$.  These transforms arise from converting the fermion-fermion or fermion-antifermion two-body interaction Hamiltonian from momentum space to coordinate space.  The first transform shown above is for the usual Coulomb interaction.  (Some methods for performing the Coulomb transform are shown in Appendix A.)  The remaining transforms were evaluated in Ref.~\onlinecite{Berestetskii82} by a series of tricks.  For Eq.~\eqref{FTb} the method is the simple expedient of writing $p_i$ in terms of the derivative of the exponential with respect to $x_i$:
\be \label{Ibeval}
I_b = -i \partial_i \int \frac{d^3 p}{(2 \pi)^3} e^{i \vec p \cdot \vec r} \, \frac{1}{p^2} = -i \partial_i \frac{1}{4 \pi r} = \frac{i x_i}{4 \pi r^3}
\ee
where $\partial_i \equiv \frac{\partial}{\partial x^i}$.  For Eq.~\eqref{FTc} they used the exponential derivative to take care of one momentum factor and a derivative of $1/p^2$ with respect to $p_j$ to represent the other.  The $p_j$ derivative was moved to the exponential by an integration by parts:
\be
I_c = -i \partial_i \int \frac{d^3 p}{(2 \pi)^3} e^{i \vec p \cdot \vec r} \left ( \frac{-\partial}{2 \partial p_j} \right ) \frac{1}{p^2} =  -i \partial_i \int \frac{d^3 p}{(2 \pi)^3} \left ( \frac{\partial}{2 \partial p_j} e^{i \vec p \cdot \vec r} \right ) \frac{1}{p^2} = -i \partial_i \left ( \frac{i x_j}{2} \frac{1}{4 \pi r} \right ) = \frac{1}{8 \pi r} \left (\delta_{i j} - \hat x_i \hat x_j \right ) .
\ee
The final transform Eq.~\eqref{FTd} is more subtle.  Berestetskii {\it et al.} used the derivative trick of Eq.~\eqref{Ibeval} twice to write
\be \label{FTd.2}
I_d = - \partial_i \partial_j \frac{1}{4 \pi r} .
\ee
They performed the derivatives in Eq.~\eqref{FTd.2} in the usual way for $r>0$ to obtain the non-delta function term of Eq.~\eqref{FTd}.  They noted that angular averaging amounts to $\partial_i \partial_j \rightarrow \frac{1}{3} \delta_{i j} \partial^2$, which with the point-source Poisson equation
\be \label{Poisson}
- \partial^2 \frac{1}{4 \pi r} = \delta(\vec r\,)
\ee
implies that there must be a term involving the three-dimensional delta function $\delta(\vec r\,)$ on the right hand side of Eq.~\eqref{FTd}.  They made the implicit assumption that the delta function would only appear in a spherically symmetric way, and noted that the angular average of $\delta_{i j}-3 \hat x_i \hat x_j$ vanishes.  With these assumptions, they obtained Eq.~\eqref{FTd}.  We will have more to say about all of these identities, especially the last, in the remainder of this work.

We will present a systematic procedure for obtaining transforms of the sort sampled in Eqs.~\eqref{FT}.  We will develop a procedure for finding transforms of the form
\be \label{FTC}
I_{n; i_1 \dots i_L}(\vec r\,) = \int \frac{d^3 p}{(2 \pi)^3} e^{i \vec p \cdot \vec r} p^n \hat p_{i_1} \cdots \hat p_{i_L}
\ee
when the transform exists.  Equivalently we will evaluate the transforms
\be \label{FTY}
I_{n \ell m}(\vec r\,) = \int \frac{d^3 p}{(2 \pi)^3} e^{i \vec p \cdot \vec r} p^n Y_\ell^m(\hat p).
\ee
The equivalence is due to the fact that any angular function, such as $\hat p_{i_1} \cdots \hat p_{i_L}$, can be written in terms of spherical harmonics $Y_\ell^m(\hat p)$.  Our method involves expressing the exponential as a Rayleigh expansion \cite{Watson22a,Morse53a,Arfken01a}
\be \label{Rayleigh}
e^{i \vec p \cdot \vec r\,} = \sum_{\ell=0}^\infty \sum_{m=-\ell}^\ell (2\ell+1) i^\ell j_\ell(p r) P_\ell(\hat p \cdot \hat r) ,
\ee
where the $j_\ell(x)$ are spherical Bessel functions and the $P_\ell(x)$ are Legendre polynomials.  We use the addition theorem for spherical harmonics \cite{Arfken01b}
\be
P_\ell(\hat p \cdot \hat r) = \frac{4 \pi}{2 \ell+1} \sum_{m=-\ell}^\ell Y_\ell^m(\hat r) Y_\ell^{m *}(\hat p)
\ee
to factor the angular dependence of $P_\ell(\hat p \cdot \hat r)$ into parts involving the angles of $\hat r$ and $\hat p$ alone.  We substitute Eq.~\eqref{Rayleigh} into Eq.~\eqref{FTY} and use the orthonormality of spherical harmonics \cite{note1}
\be
\int d \Omega_p Y_\ell^{m *} (\hat p) Y_{\ell'}^{m'} (\hat p) = \delta_{\ell \ell'} \delta_{m m'}
\ee
to write the transform as
\be
I_{n \ell m}(\vec r\,) = \frac{i^\ell}{2 \pi^2} Y_\ell^m(\hat r) \int_0^\infty dp \, p^{n+2} j_\ell(p r) .
\ee
The problem has been reduced to the evaluation of the integrals
\be \label{Rdef}
R_{n \ell}(r) \equiv \int_0^\infty dp \, p^{n+2} j_\ell(p r)
\ee
for the relevant values of $n$ and $\ell$.

Transforms of the form shown in Eq.~\eqref{FTC} can be easily expressed in terms of spatial derivatives, for instance:
\be
I_{L-2; i_1 \dots i_L}(\vec r\,) = (-i)^L \partial_{i_1} \cdots \partial_{i_L} \int \frac{d^3 p}{(2 \pi)^3} e^{i \vec p \cdot \vec r} \; \frac{1}{p^2} = (-i)^L \partial_{i_1} \cdots \partial_{i_L} \left ( \frac{1}{4 \pi r} \right ) .
\ee
For $L>1$ here the derivatives cannot be taken naively near $r = 0$ and a regulation is required (note the delta function in Eq.~\eqref{FTd}).  The regulation of derivatives of $1/r$ has received significant attention recently. \cite{Frahm83,Estrada85,Weiglhofer89,Bowen94a,Estrada95,Menon99,Hnizdo00,Aguirregabiria02,Blinder03,Hu04,Gsponer07,Franklin10,Hnizdo11}  Correspondingly, in our direct evaluation of the Fourier transforms the radial integrals $R_{n \ell}(r)$ don't necessarily converge and must be regulated.

Many of the expressions appearing in this work involve Dirac delta `functions' or other singular objects.  We consider these objects to be distributions, or generalized functions, and not functions in the usual sense. \cite{Lighthill58,Gagnon70,Challifour72,Skinner89,Reddy98}  While the values of these quantities might not be well-defined at particular points, their integrals when multiplied by an appropriate class of test functions are always well-defined and calculable.  A procedure for verifying the correctness of formulas involving generalized functions by integration with test functions is described in Appendix~B.

This article is organized as follows.  In Section~\ref{sec2} we describe the evaluation of the integrals $R_{n \ell}(r)$.  For some values of $n$, $\ell$ these integrals converge in the strict sense; for some they diverge but useful finite values can be obtained by regularization or otherwise; and for others the integrals are proportional to the delta function $\delta(r)$.  In Section~\ref{sec3} we give results for the Fourier transforms of the forms $I_{n \ell m}(\vec r\,)$ and $I_{n; i_1 \cdots i_L}(\vec r\,)$ defined above.  In Section~\ref{sec4} we describe several examples and applications, and derive a number of identities involving multiple derivatives of the functions $1/r$, $1/r^2$, and $\delta(\vec r\,)$.  In Appendix~\ref{appendix1} we show how to perform the Fourier transform for the Coulomb potential.  Finally, in Appendix~\ref{appendix2} we confirm a number of our results for generalized functions by integration with test functions.

%%%%%%%%%%%%%%%%%%%%%%%%%%%%%%%%%%%%%%%%%%%%%%%%%%%

\section{Integrals involving spherical Bessel functions}
\label{sec2}

We need values for the integrals $R_{n \ell}(r)$ defined in Eq.~\eqref{Rdef}.  For convergent integrals we can change the variable of integration to obtain $R_{n \ell}(r) = \chi_{n \ell}/r^{n+3}$ where
\be \label{chi.def}
\chi_{n \ell} = \int_0^\infty dx \, x^{n+2} j_\ell(x) .
\ee
In order to know what values of $n$ and $\ell$ lead to convergent integrals, and in order to get values, we will need to look carefully at the spherical Bessel functions. \cite{Arfken01c,Olver10} The basic definition of $j_\ell(x)$ is in terms of the usual Bessel function (of the first kind):
\be \label{j.def}
j_\ell(x) = \sqrt{\frac{\pi}{2x}} J_{\ell+1/2}(x) ,
\ee
which is actually defined for complex values of $\ell$.  A power series expansion for $j_\ell(x)$ is
\be \label{j.series}
j_\ell(x) = x^\ell \sum_{k=0}^\infty \frac{(-1)^k}{k! (2k+2 \ell+1)!!} \left ( \frac{x^2}{2} \right )^k .
\ee
We will need only integral values $\ell=0, 1, \cdots$, in which case
\be \label{j.Rayleigh}
j_\ell(x) = (-x)^\ell \left ( \frac{1}{x} \frac{d}{dx} \right )^\ell \frac{\sin x}{x} .
\ee
The first few spherical Bessel functions are:
\bse
\bearray
j_0(x) &=& \frac{\sin x}{x} , \\
j_1(x) &=& \frac{\sin x}{x^2} - \frac{\cos x}{x} , \\
j_2(x) &=& \left ( \frac{3}{x^3} - \frac{1}{x} \right ) \sin x - 3 \frac{\cos x}{x^2} .
\eearray
\ese
We see that they are relatively simple functions with oscillatory behavior that fall off for large $x$.  The small $x$ behavior is less easy to see from the explicit forms, but from the power series expression in Eq.~\eqref{j.series} it follows that
\be \label{small.x}
j_\ell(x) = \frac{x^\ell}{(2 \ell+1)!!}+O(x^{\ell+2})
\ee
for small $x$.  The large $x$ asymptotic behavior is given by
\be
j_\ell(x) \approx \frac{\sin (x-\pi \ell/2)}{x} .
\ee
It follows that in order for the integral $\chi_{n \ell}$ to exist, it must be that $-1<n+2+\ell$ lest there be a divergence at small values of $x$, and also $n<-1$ to avoid a divergence at large values of $x$.  We note that only the first and third among the transforms of Eqs.~\eqref{FT} lead to convergent radial integrals.  The tables give the value \cite{Gradshteyn80a}
\be \label{chi.integral}
\chi_{n \ell} = 2^{n+1} \sqrt{\pi} \, \frac{\Gamma \left ( \frac{\ell+3+n}{2} \right )}{\Gamma \left ( \frac{\ell-n}{2} \right )}
\ee
for the integral \eqref{chi.def} as long as $n$ is in the range of convergence: $-(\ell+3)<n<-1$ (for the case with $n$ and $\ell$ real). 

\begin{figure}
\includegraphics{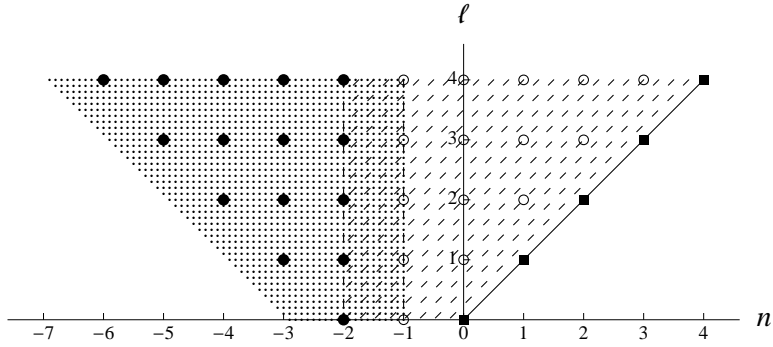}
\caption{\label{fig1} Values of $n$ and $\ell$ for which $R_{n \ell}(r) = \int_0^\infty dp \, p^{n+2} j_\ell(p r)$ is defined.  In the region on the left, with $-(\ell+3)< n <-1$ (indicated by small dots), $R_{n \ell}(r)$ and $\chi_{n \ell}=r^{n+3} R_{n \ell}(r)=\int_0^\infty dx \, x^{n+2} j_\ell(x)$ are actually convergent.  The solid dots show integral values of $n$, $\ell$ in this region.  In the region on the right, with $-2 < n < \ell$ (indicated by dashed lines),  $R_{n \ell}(r)$ and $\chi_{n \ell}$ can be given values either by extending the domain of Eq.~\eqref{chi.integral}, by reciprocity Eq.~\eqref{reciprocity}, or equivalently, by introduction of a cut-off as in Eq.~\eqref{cutoff.integral}.  Circles show integral values of $n$, $\ell$ in this region.  On the line $\ell=n$,  $R_{n \ell}(r)$ is proportional to the delta function $\delta(r)$ as in Eq.~\eqref{R.integral}.  Integral values on this line are represented by squares.}
\end{figure}

Transforms \eqref{FTb} and \eqref{FTd} have $n=-1$ and $n=0$ respectively, which are outside the range of convergence for our radial integral, so we will have to extend that range.  This extension can be done both formally and physically--fortunately, the various approaches are consistent with one another.  As a formal method of extension we can simply allow Eq.~\eqref{chi.integral} to {\it define\/} the value of $\chi_{n \ell}$ for values of $n$ and $\ell$ for which it would otherwise not be defined.  This is the same approach that is used to define values for the divergent integrals of dimensional regularization. \cite{Ashmore72,Bollini72,tHooft72,Hans83,Olness11}  This procedure allows us to extend the range of $n$ to $-(\ell+3)<n<\ell$; when $n=\ell$ the value given by the formula Eq.~\eqref{chi.integral} vanishes and hides the important fact that $R_{\ell \ell}(r)$ contains a delta function singularity at $r=0$.  The values of $n$ and $\ell$ for which integral $R_{n \ell}(r)$ can be defined in this way are represented in Fig.~\ref{fig1}.

%%%%%%
%%%%%%

Another formal approach to the extension of the allowed range of $n$ makes use of the completeness relations for spherical Bessel functions \cite{Watson22b,Morse53b,Messiah61}
\be \label{SB.completeness}
\delta(r-a) = \frac{2 r^2}{\pi} \int_0^\infty dp \, p^2 j_\ell(p r) j_\ell(p a) .
\ee
An interesting derivation of Eq.~\eqref{SB.completeness} based on the Rayleigh expansion has been given by Ugin\v{c}ius. \cite{Ugincius72}  Starting from Eq.~\eqref{SB.completeness} we divide by $a^{n+1}$ and integrate over $a$:
\be
\frac{1}{r^{n+1}} = \int_0^\infty \frac{da}{a^{n+1}} \delta(r-a) = \frac{2 r^2}{\pi} \int_0^\infty dp \, p^{n+2} j_\ell(p r) \int_0^\infty \frac{da}{a^{n+1}} \frac{1}{p^n} j_\ell(p a) .
\ee
Changing variables now to $z=p r$, $x=p a$ we obtain the reciprocity relation
\be \label{reciprocity}
\frac{\pi}{2} = \int_0^\infty dz \, z^{n+2} j_\ell(z) \int_0^\infty dx \, x^{- (n+1)}  j_\ell(x)
\ee
Relation \eqref{reciprocity} is of a formal nature as the range of convergence $-(\ell+3)<n<-1$ of the $z$ integral only overlaps with the range $-2<n<\ell$ of the $x$ integral in the region $-2<n<-1$.  Nevertheless, if we use Eq.~\eqref{chi.integral} to perform the $x$ integral in its allowed range, we find for the $z$ integral the result
\be
\int_0^\infty dz \, z^{n+2} j_\ell(z) = \frac{\pi}{2 \chi_{-(n+3), \ell}} = 2^{n+1} \sqrt\pi \, \frac{\Gamma \left ( \frac{\ell+3+n}{2} \right )}{\Gamma \left ( \frac{\ell-n}{2} \right )} = \chi_{n \ell} ,
\ee
which agrees with Eq.~\eqref{chi.integral}.  This allows us to extend the range of $n$ for which $\chi_{n \ell}$ is defined to the whole region $-(\ell+3)<n<\ell$.

As a physical regularization we make use of the expectation that the infinite extent of the range of integration is an idealization and will, in one way or another, be cut off for large values.  We model this cut-off by a mathematically expedient exponential decrease, and generalize $R_{n \ell}(r)$ to \cite{Gradshteyn80b}
\be \label{cutoff.int}
\int_0^\infty dp \, e^{- \lambda p} p^{n+2} j_\ell(p r) = \frac{\sqrt \pi r^\ell}{2^{\ell+1}} \frac{\Gamma \left ( \ell+3+n \right )}{\Gamma \left ( \ell+\frac{3}{2} \right )} \frac{_2F_1 \left ( \frac{\ell+3+n}{2}, \frac{\ell+4+n}{2}; \ell+\frac{3}{2} ; -\frac{r^2}{\lambda^2} \right )}{\lambda^{\ell+3+n}}
\ee
for positive $\lambda$.  In the limit $\lambda \rightarrow 0$ the integral \eqref{cutoff.int} becomes
\be \label{cutoff.integral}
\lim_{\lambda \rightarrow 0^+} \int_0^\infty dp \, e^{- \lambda p} p^{n+2} j_\ell(p r) = \frac{2^{n+1} \, \sqrt \pi}{r^{n+3}} \frac{\Gamma \left ( \frac{\ell+3+n}{2} \right )}{\Gamma \left ( \frac{\ell-n}{2} \right )} = \frac{\chi_{n \ell}}{r^{n+3}},
\ee
consistent with Eq.~\eqref{chi.integral} but convergent for all $-(\ell+3) < n$.

\begin{figure}
\includegraphics{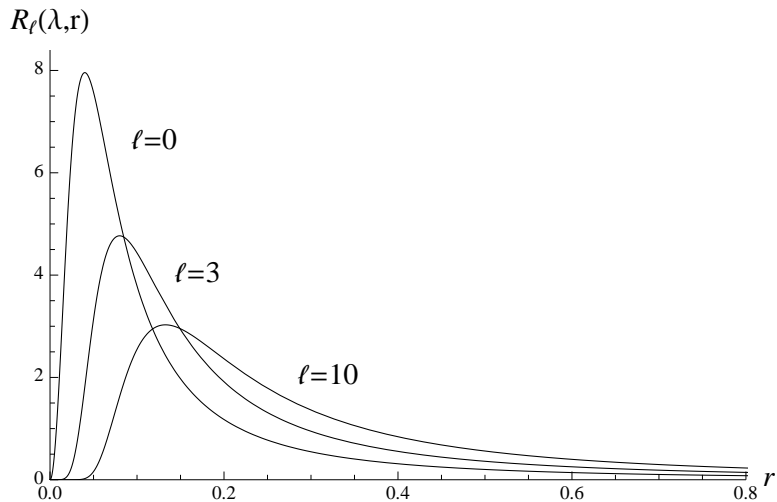}
\caption{\label{fig2} The functions $R_\ell(\lambda, r)$ that form representations of the delta function $\delta(r)$, shown for $\ell=0$, $\ell=3$, and $\ell=10$ with $\lambda=0.04$.  These functions all have unit area and a single maximum whose height is $O(1/\lambda)$, width is $O(\lambda)$, and position is a distance of $O(\lambda)$ to the right of the origin.}
\end{figure}

We will also need useful values for the integrals $R_{n \ell}(r)$ of Eq.~\eqref{Rdef} with $n=\ell$.  As a simple first example we work out the case with $\ell=0$.  Starting from Eq.~\eqref{SB.completeness} in the limit $a \rightarrow 0$, we find
\be
\delta(r) = \frac{2 r^2}{\pi} \int_0^\infty dp \, p^2 j_0(p r) .
\ee
This result has been noted in Ref.~\onlinecite{Mehrem11}.  It is a consequence of the usual Fourier representation for the delta function
\be
\delta(\vec r\,) = \int \frac{d^3 p}{(2 \pi)^3} e^{i \vec p \cdot \vec r} = \frac{1}{4 \pi^2} \int_0^\infty dp \, p^2 \int_{-1}^1 du \, e^{i p r u} = \frac{1}{2 \pi^2} \int_0^\infty dp \, p^2 \frac{\sin(p r)}{p r} = \frac{1}{2 \pi^2} \int_0^\infty dp \, p^2 j_0(p r)
\ee
(where $u=\cos \theta_p$ as in Appendix A) after use of the delta function identity $\delta(\vec r\,) = \delta(r)/(4 \pi r^2)$ (see Appendix~B for a proof).  In order to obtain the needed generalization, we rewrite Eq.~\eqref{SB.completeness} remembering that the integral vanishes unless $r=a$:
\be \label{SB.completeness.prime}
\delta(r-a) = \frac{2 r^{\ell+2}}{\pi a^\ell} \int_0^\infty dp \, p^2 j_\ell(p r) j_\ell(p a)
\ee
and take the $a \rightarrow 0$ limit using Eq.~\eqref{small.x}: \cite{note2}
\be
\delta(r) = \frac{2 r^{\ell+2}}{\pi (2 \ell+1)!!} \int_0^\infty dp \, p^{\ell+2} j_\ell(p r) ,
\ee
or
\be \label{R.integral}
R_{\ell \ell}(r)  = \int_0^\infty dp \, p^{\ell+2} j_\ell(p r) = \frac{\pi (2 \ell+1)!!}{2 r^{\ell+2}} \delta(r) = \frac{2 \pi^2 (2\ell+1)!!}{r^\ell} \delta(\vec r\,) .
\ee
We now bolster this formal derivation with a physical one based on the inclusion of a cut-off.  The cut-off version of $R_{\ell \ell}(r)$ is the integral
\be
R_\ell(\lambda; r) \equiv \frac{2 r^{\ell+2}}{\pi (2 \ell+1)!!} \int_0^\infty dp \, e^{-\lambda p} p^{\ell+2} j_\ell(p r)
=\frac{2^{\ell+2} (\ell+1)!}{\pi (2 \ell+1)!!} \frac{\lambda r^{2 \ell+2}}{(r^2+\lambda^2)^{\ell+2}} .
\ee
(We have included a pre-factor for convenience.)  Then $R_\ell(\lambda; r)$ is a representation of the delta function.  It is properly normalized:
\be
\int_0^\infty dr \, R_\ell(\lambda; r) = 1
\ee
for all positive values of $\lambda$. \cite{note3} The functions $R_\ell(\lambda; r)$ have the general shape shown in Fig.~\ref{fig2}. \cite{note4}  For each value of $\ell$, $R_\ell(\lambda; r)$ has a single maximum of width $O(\lambda)$, height $O(1/\lambda)$, and position a distance of $O(\lambda)$ to the right of $r=0$.  The functions $R_\ell(\lambda; r)$ satisfy the sifting property
\be
\lim_{\lambda \rightarrow 0} \int_0^\infty dr \, f(r) R_\ell(\lambda; r) = f(0)
\ee
for all continuous functions $f(r)$ that vanish sufficiently rapidly at large $r$, and so form representations of $\delta(r)$ for each value of $\ell$ (including non-integral values of $\ell$ as long as $-3/2<\ell$).  We notice that $\chi_{n \ell}=0$ when $n=\ell$ from Eq.~\eqref{chi.integral} is consistent with $\chi_{\ell \ell}=R_{\ell \ell}(1) \propto \delta(1)=0$ from Eq.~\eqref{R.integral}.

%%%%%%
%%%%%%

%%%%%%%%%%%%%%%%%%%%%%%%%%%%%%%%%%%%%%%%%%%%%%%%%%%

\section{Results for the Fourier transforms}
\label{sec3}

We can now present results for the three-dimensional Fourier transforms considered in this paper.  The transforms of the form of Eq.~\eqref{FTY} that we can do are
\be \label{FTY.result1}
I_{n \ell m}(\vec r\,) = \int \frac{d^3 p}{(2 \pi)^3} e^{i \vec p \cdot \vec r} p^n Y_\ell^m(\hat p) = \frac{i^\ell}{2 \pi^2} \frac{\chi_{n \ell}}{r^{n+3}} \, Y_\ell^m(\hat r) ,
\ee
when $-(\ell+3) < n < \ell$, and
\be \label{FTY.result2}
I_{\ell \ell m}(\vec r\,) = \int \frac{d^3 p}{(2 \pi)^3} e^{i \vec p \cdot \vec r} \, p^\ell \, Y_\ell^m(\hat p) = i^\ell \frac{(2\ell+1)!!}{4 \pi r^{\ell+2}} \delta(r) Y_\ell^m(\hat r) = i^\ell \frac{(2\ell+1)!!}{r^\ell} \delta(\vec r\,) Y_\ell^m(\hat r) ,
\ee
which is the form appropriate when $n=\ell$.  Results for these transforms have a simple form because each contains only a single value of angular momentum $\ell$, and so involves only a single spherical Bessel function.

A more general result for transforms like that of Eq.~\eqref{FTY.result1} can also be found that generalizes the number of spatial dimensions to $N$ and allows for arbitrary complex values of $n$. \cite{Samko78}  The general result has a form much like that of Eq.~\eqref{FTY.result1} (but generalized to $N$ dimensions) except for the exceptional lines $\ell=n-2k$ (where $k=0, 1, 2, \dots$) on which the transform contains the delta function or its derivatives, and $\ell=-n-N-2k$ (where $k=0, 1, 2, \dots$) on which the transform has a somewhat complicated form containing $\ln r$ and powers of $r$ but no delta functions.

The transforms \eqref{FTC} are more complicated than those of \eqref{FTY} because they involve multiple values of $\ell$.  We reduce Eq.~\eqref{FTC} to a sum of terms like Eq.~\eqref{FTY} by expanding each angular factor $\hat p_{i_1} \cdots \hat p_{i_L}$ in terms of spherical harmonics as
\be \label{C.in.terms.of.Y}
\hat p_{i_1} \cdots \hat p_{i_L} = \sum_{\ell=L}^{1 \, {\rm or} \, 0} \sum_{m=-\ell}^\ell C^{\ell m}_{i_1 \cdots i_L} Y_\ell^m(\hat p) ,
\ee
where $\ell$ in the sum takes the values $L$, $L-2$, \dots, $1$ or $0$ (depending on whether $L$ is odd or even).  This expansion can be understood by remembering that each factor of $\hat p_i$ has $\ell=1$, as follows from the fact that each component of $\hat p$ can be expressed linearly in terms of $Y_1^m(\hat p)$, and the addition law of angular momenta implies that the maximum value of $\ell$ obtained from $L$ factors of unit angular momentum is $L$.  Parity tells us that only odd or only even values of $\ell$ contribute--as $\hat p_{i_1} \cdots \hat p_{i_L}$ has parity $(-1)^L$ and $Y_{\ell}^m(\hat p)$ has parity $(-1)^\ell$, and all terms of Eq.~\eqref{C.in.terms.of.Y} must have the same parity.  Explicit values of the $C^{\ell m}_{i_1 \cdots i_L}$ coefficients can be found using the orthogonality of the spherical harmonics:
\be \label{C.coefs}
C^{\ell m}_{i_1 \cdots i_L} = \int d \Omega_p \, Y_\ell^{m *} (\hat p) \, \hat p_{i_1} \cdots \hat p_{i_L} .
\ee

We turn now to a study of the angular momentum decomposition of $\hat p_{i_1} \cdots \hat p_{i_L}$.  From our previous discussion it is clear that $\hat p_{i_1} \cdots \hat p_{i_L}$ contains contributions of angular momenta $L$, $L-2$, etc., down to $1$ or $0$.  We define $\left ( \hat p_{i_1} \cdots \hat p_{i_L} \right )^L_\ell$ to be that part of $\hat p_{i_1} \cdots \hat p_{i_L}$ of angular momentum $\ell$, {\it i.e.\,}
\be \label{angular.decomposition}
\left ( \hat p_{i_1} \cdots \hat p_{i_L} \right )^L_\ell \equiv \sum_{m=-\ell}^\ell C^{\ell m}_{i_1 \cdots i_L} Y_\ell^m(\hat p) .
\ee
An explicit expression for $\left ( \hat p_{i_1} \cdots \hat p_{i_L} \right )^L_\ell$ can be obtained by using Eq.~\eqref{C.coefs} in Eq.~\eqref{angular.decomposition} and the addition theorem for spherical harmonics:
\be \label{angular.decomposition.2}
\left ( \hat p_{i_1} \cdots \hat p_{i_L} \right )^L_\ell = \int d \Omega_{p'} \, \hat p'_{i_1} \cdots \hat p'_{i_L} \sum_{m=-\ell}^\ell Y_\ell^{m *}(\hat p') Y_\ell^m(\hat p) =  (2\ell+1) \int \frac{d \Omega_{p'}}{4 \pi} \, \hat p'_{i_1} \cdots \hat p'_{i_L} \, P_\ell(\hat p' \cdot \hat p) .
\ee
The method is to write out $P_\ell(\hat p' \cdot \hat p)$ as a polynomial of order $\ell$ and perform the angular integral using \cite{Bowen94b}
\be \label{angular.average}
\int \frac{d \Omega}{4 \pi} \, \hat x_{i_1} \cdots \hat x_{i_N} = \frac{\delta_{N,{\rm even}}}{(N+1)!!} \left ( \delta_{i_1 i_2} \delta_{i_3 i_4} \cdots \delta_{i_{N-1} i_N} + {\rm perms} \right )_{(N-1)!! \; {\rm terms}} \, .
\ee
(The integral vanishes for odd $N$, and the subscript indicates that there are a total of $(N-1)!!$ terms in the sum over permutations.)  Using Eqs.~\eqref{angular.decomposition.2} and \eqref{angular.average}, it is not difficult to obtain $\left ( \hat p_{i_1} \cdots \hat p_{i_L} \right )^L_\ell$ for $\ell=0$ and $\ell=1$ because the corresponding Legendre polynomials are so simple.  (The first two Legendre polynomials are $P_0(\hat p' \cdot \hat p)=1$ and $P_1(\hat p' \cdot \hat p) = \hat p' \cdot \hat p = p'_j p_j$ with an understood sum over $j$ from 1 to 3.)  We find that
\bse \label{inits}
\bearray
\left ( \hat p_{i_1} \cdots \hat p_{i_L} \right )^L_0 &=& \int \frac{d \Omega_{p'}}{4 \pi} \, \hat p'_{i_1} \cdots \hat p'_{i_L} = \frac{\delta_{L,{\rm even}}}{(L+1)!!} \left ( \delta_{i_1 i_2} \cdots \delta_{i_{L-1} i_L} + {\rm perms} \right )_{(L-1)!! \; {\rm terms}} \; , \\
\left ( \hat p_{i_1} \cdots \hat p_{i_L} \right )^L_1 &=& 3 \int \frac{d \Omega_{p'}}{4 \pi} \, \hat p'_{i_1} \cdots \hat p'_{i_L} \hat p'_j \, \hat p_j  = 3 \left ( \hat p_{i_1} \cdots \hat p_{i_L} \hat p_j \right )^{L+1}_0 \, \hat p_j \cr
&\hbox{}& \quad = \frac{3 \delta_{L,{\rm odd}}}{(L+2)!!} \left ( \hat p_{i_1} \left ( \delta_{i_2 i_3} \cdots \delta_{i_{L-1} i_L} + {\rm perms} \right )_{(L-2)!! \; {\rm terms}} + {\rm perms} \right )_{L \; {\rm terms}} \, .
\eearray
\ese
This direct approach becomes increasingly cumbersome for higher values of $\ell$.  However, the results obtained in Eqs.~\eqref{inits} can serve as initial values for an inductive proof of a general expression for $\left ( \hat p_{i_1} \cdots \hat p_{i_L} \right )^L_\ell$. \cite{Adkinsxx}

The most convenient way to obtain $\left ( \hat p_{i_1} \cdots \hat p_{i_L} \right )^L_\ell$, at least for small $L$, is by straightforward construction.  We explain the method here and give several examples.  We note first that the quantity $\left ( \hat p_{i_1} \cdots \hat p_{i_L} \right )^L_\ell$ is symmetric in all indices.  Furthermore, the part with maximal angular momentum, $\left ( \hat p_{i_1} \cdots \hat p_{i_L} \right )^L_L$, is traceless:
\be
\left ( \hat p_{i_1} \cdots \hat p_{i_L} \right )^L_L \delta_{i_{L-1}, i_L} = \left ( \hat p_{i_1} \cdots \hat p_{i_{L-2}} \right )^{L-2}_L = 0 ,
\ee
because an object of angular momentum $L$ cannot be constructed from a combination of only $L-2$ objects each having angular momentum one.  The condition of tracelessness allows us to construct $\left ( \hat p_{i_1} \cdots \hat p_{i_L} \right )^L_L$ relatively easily for each particular value of $L$.  The $L=2$ result is immediate:
\be
\left ( \hat p_{i_1} \hat p_{i_2} \right )^2_2 = \hat p_{i_1} \hat p_{i_2} - \frac{1}{3} \delta_{i_1 i_2} .
\ee
The remainder is the $\ell=0$ contribution:
\be
\left ( \hat p_{i_1} \hat p_{i_2} \right )^2_0 = \frac{1}{3} \delta_{i_1 i_2} ,
\ee
so that
\be
\hat p_{i_1} \hat p_{i_2} = \left ( \hat p_{i_1} \hat p_{i_2} \right )^2_2 + \left ( \hat p_{i_1} \hat p_{i_2} \right )^2_0 .
\ee

For $L=3$ we start with $\hat p_{i_1} \hat p_{i_2} \hat p_{i_3}$ and add a symmetric term with two fewer momentum unit vectors but the same three indices with an initially undetermined coefficient:
\be
\left ( \hat p_{i_1} \hat p_{i_2} \hat p_{i_3} \right )^3_3 =  \hat p_{i_1} \hat p_{i_2} \hat p_{i_3} + A \left ( \hat p_{i_1} \delta_{i_2 i_3} + \hat p_{i_2} \delta_{i_3 i_1} + \hat p_{i_3} \delta_{i_1 i_2} \right ) \, .
\ee
Contraction over $i_2$ and $i_3$ yields
\be
0 = \left ( \hat p_{i_1} \hat p_{i_2} \hat p_{i_3} \right )^3_3 \delta_{i_2 i_3} = \hat p_{i_1} + A \left ( 5 \hat p_{i_1} \right ) ,
\ee
so that $A=-1/5$.  So for $L=3$ we find
\be
\left ( \hat p_{i_1} \hat p_{i_2} \hat p_{i_3} \right )^3_3 =  \hat p_{i_1} \hat p_{i_2} \hat p_{i_3} - \frac{1}{5} \left ( \hat p_{i_1} \delta_{i_2 i_3} + \hat p_{i_2} \delta_{i_3 i_1} + \hat p_{i_3} \delta_{i_1 i_2} \right )
\ee
as the maximal angular momentum term, and the remainder is the $\ell=1$ contribution:
\be
\left ( \hat p_{i_1} \hat p_{i_2} \hat p_{i_3} \right )^3_1 = \frac{1}{5} \left ( \hat p_{i_1} \delta_{i_2 i_3} + \hat p_{i_2} \delta_{i_3 i_1} + \hat p_{i_3} \delta_{i_1 i_2} \right ) .
\ee
By construction, we have the angular momentum decomposition:
\be
\hat p_{i_1} \hat p_{i_2} \hat p_{i_3} = \left ( \hat p_{i_1} \hat p_{i_2} \hat p_{i_3} \right )^3_3 + \left ( \hat p_{i_1} \hat p_{i_2} \hat p_{i_3} \right )^3_1 .
\ee

The term with four momentum unit vectors can be decomposed in an analogous way:
\be
\left ( \hat p_{i_1} \hat p_{i_2} \hat p_{i_3} \hat p_{i_4} \right )^4_4 = \hat p_{i_1} \hat p_{i_2} \hat p_{i_3} \hat p_{i_4} - \frac{1}{7} \left ( \hat p_{i_1} \hat p_{i_2} \delta_{i_3 i_4} + {\rm perms} \right )_{\rm{6 \; terms}} + \frac{1}{35} \left ( \delta_{i_1 i_2} \delta_{i_3 i_4} + {\rm perms} \right )_{\rm{3 \; terms}} .
\ee
The $\ell=2$ term of $\hat p_{i_1} \hat p_{i_2} \hat p_{i_3} \hat p_{i_4}$ comes from subtracting the appropriate traces from the parts of $\hat p_{i_1} \hat p_{i_2} \hat p_{i_3} \hat p_{i_4} - \left ( \hat p_{i_1} \hat p_{i_2} \hat p_{i_3} \hat p_{i_4} \right )^4_4$ that are quadratic in momentum unit vectors:
\be
\left ( \hat p_{i_1} \hat p_{i_2} \hat p_{i_3} \hat p_{i_4} \right )^4_2 = \frac{1}{7} \big ( \left ( \hat p_{i_1} \hat p_{i_2} \right )^2_2 \delta_{i_3 i_4}  + {\rm perms} \big )_{\rm 6 \; terms} .
\ee
The $\ell=0$ part is what remains:
\be
\left ( \hat p_{i_1} \hat p_{i_2} \hat p_{i_3} \hat p_{i_4} \right )^4_0 =  \frac{1}{15} \left ( \delta_{i_1 i_2} \delta_{i_3 i_4} + {\rm perms} \right )_{\rm{3 \; terms}} .
\ee

As a final example, we give the results for $L=5$:
\bse
\bearray
\left ( \hat p_{i_1} \hat p_{i_2} \hat p_{i_3} \hat p_{i_4} \hat p_{i_5} \right )^5_5 &=& \hat p_{i_1} \hat p_{i_2} \hat p_{i_3} \hat p_{i_4} \hat p_{i_5} - \frac{1}{9} \left ( \hat p_{i_1} \hat p_{i_2} \hat p_{i_3} \delta_{i_4 i_5} + {\rm perms} \right )_{\rm{10 \; terms}} + \frac{1}{63} \left ( \hat p_{i_1} \delta_{i_2 i_3} \delta_{i_4 i_5} + {\rm perms} \right )_{\rm{15 \; terms}} , \\
\left ( \hat p_{i_1} \hat p_{i_2} \hat p_{i_3} \hat p_{i_4} \hat p_{i_5} \right )^5_3 &=& \frac{1}{9} \big ( \left ( \hat p_{i_1} \hat p_{i_2} \hat p_{i_3} \right )^3_3 \delta_{i_4 i_5} + {\rm perms} \big )_{\rm{10 \; terms}} , \\
\left ( \hat p_{i_1} \hat p_{i_2} \hat p_{i_3} \hat p_{i_4} \hat p_{i_5} \right )^5_1 &=& \frac{1}{35} \big ( \hat p_{i_1} \delta_{i_2 i_3} \delta_{i_4 i_5} + {\rm perms} \big )_{\rm{15 \; terms}} .
\eearray
\ese

We note that angular decomposition works for vectors with non-unit length just as well as for unit vectors.  One simply has that
\be
\left ( p_{i_1} \cdots p_{i_L} \right )^L_\ell = p^L \left (\hat p_{i_1} \cdots \hat p_{i_L} \right )^L_\ell .
\ee
So, for instance,
\be
p_{i_1} p_{i_2} p_{i_3} = \left (p_{i_1} p_{i_2} p_{i_3} \right )^3_3 + \left (p_{i_1} p_{i_2} p_{i_3} \right )^3_1
\ee
where
\be
\left (p_{i_1} p_{i_2} p_{i_3} \right )^3_3 = p_{i_1} p_{i_2} p_{i_3} - \frac{p^2}{5} \left ( p_{i_1} \delta_{i_2 i_3} + p_{i_2} \delta_{i_3 i_1} + p_{i_3} \delta_{i_1 i_2} \right )
\ee
and
\be
\left (p_{i_1} p_{i_2} p_{i_3} \right )^3_1 = \frac{p^2}{5} \left ( p_{i_1} \delta_{i_2 i_3} + p_{i_2} \delta_{i_3 i_1} + p_{i_3} \delta_{i_1 i_2} \right ) .
\ee

We are now ready to obtain results for the transforms $I_{n; i_1 \cdots i_L}(\vec r\,)$ of Eq.~\eqref{FTC}.  First, for the component $\left ( \hat p_{i_1} \cdots \hat p_{i_L} \right )^L_\ell$ of angular momentum $\ell$, one has
\be \label{FTC.result1}
\int \frac{d^3 p}{(2 \pi)^3} e^{i \vec p \cdot \vec r} p^n \left ( \hat p_{i_1} \cdots \hat p_{i_L} \right )^L_\ell = \frac{i^\ell}{2 \pi^2} \frac{\chi_{n \ell}}{r^{n+3}} \, \left ( \hat x_{i_1} \cdots \hat x_{i_L} \right )^L_\ell ,
\ee
valid when $-(\ell+3) < n < \ell$.  The result \eqref{FTC.result1} comes from Eq.~\eqref{FTY.result1}, the expression \eqref{angular.decomposition} giving $\left ( \hat p_{i_1} \cdots \hat p_{i_L} \right )^L_\ell$ in terms of $Y_\ell^m(\hat p)$, and an analogous expresssion for $\left ( \hat x_{i_1} \cdots \hat x_{i_L} \right )^L_\ell$ in terms of $Y_\ell^m(\hat r)$:
\be \label{angular.decomposition.r}
\left ( \hat x_{i_1} \cdots \hat x_{i_L} \right )^L_\ell \equiv \sum_{m=-\ell}^\ell C^{\ell m}_{i_1 \cdots i_L} Y_\ell^m(\hat r) .
\ee
When $n=\ell$ the transform contains a delta function: 
\be \label{FTC.result2}
\int \frac{d^3 p}{(2 \pi)^3} e^{i \vec p \cdot \vec r} \, p^\ell \, \left ( \hat p_{i_1} \cdots \hat p_{i_L} \right )^L_\ell = i^\ell \frac{(2\ell+1)!!}{r^\ell} \delta(\vec r\,) \left ( \hat x_{i_1} \cdots \hat x_{i_L} \right )^L_\ell 
\ee
by use of Eq.~\eqref{FTY.result2}.  Transforms $I_{n;i_1 \cdots i_L}(\vec r\,)$ of the function
\be
p^n \hat p_{i_1} \cdots \hat p_{i_L} = p^n \sum_\ell \left ( \hat p_{i_1} \cdots \hat p_{i_L} \right )^L_\ell
\ee
are done through use of superposition.

The Fourier transforms that we have done are summarized in Table~\ref{summary}, which shows the transform pairs $\Phi(\vec p\,)$ and $\Psi(\vec r\,)$.  The functions $\Phi(\vec p\,)$ and $\Psi(\vec r\,)$ are related by
\bse \label{FT.def}
\bearray
\Psi(\vec r\,) &=& \int \frac{d^3 p}{(2 \pi)^3} \, e^{i \vec p \cdot \vec r} \Phi(\vec p\,) \, , \\
\Phi(\vec p\,) &=& \int d^3 r \, e^{-i \vec p \cdot \vec r} \Psi(\vec r\,) \, .
\eearray
\ese
(The radial part of this transform can be expressed as a Hankel transform, \cite{Oberhettinger72,Davies78} so many additional transform pairs are also known.)

\begin{table}
\begin{center}
\caption{\label{summary} Summary of the three-dimensional Fourier transforms evaluated in this work.  The functions $\Phi(\vec p\,)$ and $\Psi(\vec r\,)$ are a transform pair as defined in Eqs.~\eqref{FT.def}.  In each of these transform pairs one can make the replacement $Y_\ell^m(\hat p) \rightarrow \left ( \hat p_{i_1} \cdots \hat p_{i_L} \right )^L_\ell$ and $Y_\ell^m(\hat r) \rightarrow \left ( \hat x_{i_1} \cdots \hat x_{i_L} \right )^L_\ell$ to obtain a different representation of the angular part of that transform.  The value of $n$ is restricted to the range $-(\ell+3) < n < \ell$.  The value of $\chi_{n \ell}$ is given in Eq.~\eqref{chi.integral}.}
\begin{ruledtabular}
\begin{tabular}{cc}
$\Phi(\vec p\,)$ & $\Psi(\vec r\,)$  \\
\hline\noalign{\smallskip}
$p^n Y_\ell^m(\hat p)$ & $\frac{i^\ell}{2 \pi^2} \frac{\chi_{n \ell}}{r^{n+3}} Y_\ell^m(\hat r)$ \\
$p^\ell Y_\ell^m(\hat p)$ & $i^\ell \frac{(2\ell+1)!!}{r^\ell} \delta(\vec r\,) Y_\ell^m(\hat r)$ \\
$\frac{1}{p^\ell} \delta(\vec p\,) Y_\ell^m(\hat p)$ & $\frac{i^\ell}{(2 \pi)^3} \frac{r^\ell}{(2\ell+1)!!} Y_\ell^m(\hat r)$ \\
$4 \pi (-i)^\ell \frac{\chi_{n \ell}}{p^{n+3}} Y_\ell^m(\hat p)$ & $r^n Y_\ell^m(\hat r)$ \\
$(-i)^\ell \frac{(2\ell+1)!!}{p^\ell} (2 \pi)^3 \delta(\vec p\,) Y_\ell^m(\hat p)$ & $r^\ell Y_\ell^m(\hat r)$ \\
$(-i)^\ell \frac{p^\ell}{(2\ell+1)!!} Y_\ell^m(\hat p)$ & $\frac{1}{r^\ell} \delta(\vec r\,) Y_\ell^m(\hat r)$ \\
\end{tabular}
\end{ruledtabular}
\end{center}
\end{table}

%%%%%%%%%%%%%%%%%%%%%%%%%%%%%%%%%%%%%%%%%%%%%%%%%%%

\section{Applications and consequences}
\label{sec4}

As first examples of the use of our Fourier transform formulas we return to the transforms of Eqs.~\eqref{FT}.  The first three are immediate applications of Eq.~\eqref{FTC.result1}.  One has for $I_a$ the result
\be
I_a = \int \frac{d^3 p}{(2 \pi)^3} e^{i \vec p \cdot \vec r} \, \frac{1}{p^2} = \frac{i^0}{2 \pi^2} \frac{\chi_{-2, 0}}{r} = \frac{1}{4 \pi r} ,
\ee
as $\chi_{-2,0} = \pi/2$ and the angular decomposition of $L$ momentum vectors when $L=0$ is $()^0_0 \equiv 1$.  Terms $I_b$ and $I_c$ can be done just as easily.  For $I_d$ we find
\be
I_d = \int \frac{d^3 p}{(2 \pi)^3} e^{i \vec p \cdot \vec r} \left \{ (\hat p_i \hat p_j)^2_0 + (\hat p_i \hat p_j)^2_2 \right \} = \delta(\vec r\,) (\hat x_i \hat x_j)^2_0 + \frac{i^2}{2 \pi^2} \frac{\chi_{0, 2}}{r^3} (\hat x_i \hat x_j)^2_2 = \frac{1}{3} \delta_{i j} \delta(\vec r \,) - \frac{3}{4 \pi r^3} \left ( \hat x_i \hat x_j - \frac{1}{3} \delta_{i j} \right )
\ee
because $\chi_{0, 2} = 3 \pi/2$.  According to Eq.~\eqref{FTd.2} we can write the transform $I_d$ as a double spatial derivative of $1/r$, which leads to
\be \label{Frahm1}
\partial_i \partial_j \frac{1}{r} = -\frac{4 \pi}{3} \delta_{i j} \delta(\vec r\,) + \frac{3}{r^3} \left ( \hat x_i \hat x_j - \frac{1}{3} \delta_{i j} \right ) .
\ee
We note that Eq.~\eqref{Frahm1} is the sum of two terms: one that comes from the usual derivative formulas and that holds when $r \ne 0$, with the other contributing only when $r=0$.  This identity was emphasized by Frahm, \cite{Frahm83} and has many applications.  Its immediate applications are for the calculation of the electric and magnetic fields for electric and magnetic dipoles. \cite{Gagnon70,Griffiths92,Leung06} The electric and magnetic dipole potentials are given in terms of the electric and magnetic dipole moments $\vec p$ and $\vec m$ as $V_e = \vec p \cdot \vec r/r^3$ and $\vec A_m = \vec m \times \vec r/r^3$.  The corresponding fields are
\be \label{electric.field}
\vec E = -\vec \nabla V_e = \hat e_i \partial_i \left ( p_j \partial_j \frac{1}{r} \right ) = -\frac{4 \pi}{3} \vec p \, \delta(\vec r\,) + \frac{3 (\vec p \cdot \vec r \,)\vec r - r^2 \vec p}{r^5}
\ee
and
\be \label{magnetic.field}
\vec B = \vec \nabla \times \vec A_m = -\vec \nabla \times \left ( \vec m \times \vec \nabla \frac{1}{r} \right ) = -\vec m \vec \nabla^2 \frac{1}{r} + \vec \nabla \left ( \vec m \cdot \vec \nabla \right ) \frac{1}{r} = \frac{8 \pi}{3} \vec m \, \delta(\vec r\,) + \frac{3 (\vec m \cdot \vec r \,)\vec r - r^2 \vec m}{r^5} ,
\ee
where $\vec \nabla$ is the gradient vector ($\nabla_i = \partial_i$) and $\hat e_i$ is the unit vector along the $i^{\rm th}$ coordinate axis.  (We have used the vector identity $\vec A \times \left ( \vec B \times \vec C \right ) = \vec B \left ( \vec A \cdot \vec C \right ) - \vec C \left ( \vec A \cdot \vec B \right )$.)  The electric and magnetic dipole fields have identical structure outside the source but are quite different at the position of the dipole.  The electric dipole is produced by separated charges of opposite sign, with the dipole moment $\vec p$ pointing from the negative charge to the positive one.  The field between the charges points from the positive charge to the negative one as represented by the negative sign on the $\vec p \delta(\vec r\,)$ term in Eq.~\eqref{electric.field}.  The magnetic dipole is produced by an infinitesimal current loop with all of the flux of the dipole flowing up through the loop in the direction of the dipole moment $\vec m$, corresponding to a positive sign on the $\vec m \delta(\vec r\,)$ term in Eq.~\eqref{magnetic.field}.  The delta function term in Eq.~\eqref{magnetic.field} is the origin of the hyperfine splitting for atomic $S$ states. \cite{Fermi30,Bethe57}  In particular, in the ground state of atomic hydrogen it leads to the $21 cm$ line \cite{Ewen51,Muller51,Griffiths82} that is important in radio astronomy.

We now turn to a number of additional Fourier transforms that give interesting results.  First we consider the transform of $\frac{1}{p^2} \left ( p_{j_1} \cdots p_{j_k} \right )^k_k$.  On the one hand, this transform can be written in terms of derivatives:
\be
\int \frac{d^3 p}{(2 \pi)^3} e^{i \vec p \cdot \vec r} \frac{1}{p^2} \left ( p_{j_1} \cdots p_{j_k} \right )^k_k = (-i)^k \left ( \partial_{j_1} \cdots \partial_{j_k} \right )^k_k \frac{1}{4 \pi r} ,
\ee
while on the other hand, we can evaluate the transform using Eq.~\eqref{FTC.result1}:
\be
\int \frac{d^3 p}{(2 \pi)^3} e^{i \vec p \cdot \vec r} \frac{1}{p^2} \left ( p_{j_1} \cdots p_{j_k} \right )^k_k = \frac{i^k}{2 \pi^2} \frac{\chi_{k-2, k}}{r^{k+1}} \left ( \hat x_{j_1} \cdots \hat x_{j_k} \right )^k_k, 
\ee
where from Eq.~\eqref{chi.integral} we obtain $\chi_{k-2, k} = \frac{\pi}{2} (2k-1)!!$.  We find then for the derivatives of $1/r$ the formula
\be \label{onebyr}
\left ( \partial_{j_1} \cdots \partial_{j_k} \right )^k_k \frac{1}{r} = \frac{(-1)^k (2k-1)!!}{r^{k+1}} \left ( \hat x_{j_1} \cdots \hat x_{j_k} \right )^k_k .
\ee
We now write this out explicitly for a few low values of $k$.  For $k=0$ the identity is trivial ($1/r=1/r$) where we recall that $()^0_0=1$ and $(-1)!! \equiv 1$.  For $k=1$ we obtain the expected result:
\be
\partial_i \frac{1}{r} = -\frac{\hat x_i}{r^2} .
\ee
For $k=2$ we obtain
\be
\left ( \partial_i \partial_j - \frac{1}{3} \delta_{i j} \partial^2 \right ) \frac{1}{r} = \frac{3}{r^3} \left ( \hat x_i \hat x_j - \frac{1}{3} \delta_{i j} \right ) ,
\ee
which, with use of the Poisson equation \eqref{Poisson}, leads again to Eq.~\eqref{Frahm1}.  The $k=3$ result can be expressed as
\be
\partial_i \partial_j \partial_k \frac{1}{r} = -\frac{15}{r^4} \left ( \hat x_i \hat x_j \hat x_k \right )^3_3 - \frac{4 \pi}{5} \big ( \partial_i \delta(\vec r\,) \delta_{j k} + \partial_j \delta(\vec r\,) \delta_{k i} + \partial_k \delta(\vec r\,) \delta_{i j} \big ) ,
\ee
an identity also obtained by Frahm. \cite{Frahm83}  We point out that although the general $k^{\rm th}$ derivative of $1/r$ contains delta functions, the particular combination $\left ( \partial_{j_1} \cdots \partial_{j_k} \right )^k_k \frac{1}{r}$ with maximal angular momentum has none.

The second new transform we consider is that of $\frac{1}{p} \left ( p_{j_1} \cdots p_{j_k} \right )^k_k$.  This transform can be represented in terms of derivatives of $1/(2 \pi^2 r^2)$ (the transform of $1/p$), or evaluated explicitly using Eq.~\eqref{FTC.result1}:
\be
(-i)^k \left ( \partial_{j_1} \cdots \partial_{j_k} \right )^k_k \frac{1}{2 \pi^2 r^2} = \int \frac{d^3 p}{(2 \pi)^3} e^{i \vec p \cdot \vec r} \frac{1}{p} \left ( p_{j_1} \cdots p_{j_k} \right )^k_k = \frac{i^k}{2 \pi^2} \frac{\chi_{k-1, k}}{r^{k+2}} \left ( \hat x_{j_1} \cdots \hat x_{j_k} \right )^k_k .
\ee
Making use of $\chi_{k-1, k} = 2^k k!$, we find that
\be \label{onebyrsq}
\left ( \partial_{j_1} \cdots \partial_{j_k} \right )^k_k \frac{1}{r^2} = \frac{(-1)^k 2^k k!}{r^{k+2}} \left ( \hat x_{j_1} \cdots \hat x_{j_k} \right )^k_k .
\ee
Again, the combination of derivatives of $1/r^2$ having maximal angular momentum gives a result without delta functions.  Relation \eqref{onebyrsq} is trivial for $k=0$.  For $k=1$ it gives
\be \label{onebyrsq.1}
\partial_i \frac{1}{r^2} = -\frac{2 \hat x_i}{r^3} ,
\ee
and for $k=2$
\be \label{onebyrsq.2}
\left ( \partial_i \partial_j - \frac{1}{3} \delta_{i j} \partial^2 \right ) \frac{1}{r^2} = \frac{8}{r^4} \left ( \hat x_i \hat x_j - \frac{1}{3} \delta_{i j} \right ) .
\ee
These formulas are easy to verify for $r>0$; the new result is that no extra delta functions appear.

The final new transform we consider is that of $\left ( p_{j_1} \cdots p_{j_k} \right )^k_k$. As before, this transform can be represented either in terms of derivatives (this time of $\delta(\vec r\,)$), or explicitly by use of Eq.~\eqref{FTC.result2}.  We find that
\be \label{delta}
(-i)^k \left ( \partial_{j_1} \cdots \partial_{j_k} \right )^k_k \delta(\vec r\,) = \int \frac{d^3 p}{(2 \pi)^3} e^{i \vec p \cdot \vec r} \left ( p_{j_1} \cdots p_{j_k} \right )^k_k = i^k \frac{(2k+1)!!}{r^k} \delta(\vec r\,) \left ( \hat x_{j_1} \cdots \hat x_{j_k} \right )^k_k .
\ee
The new delta function identities are
\be
\left ( \partial_{j_1} \cdots \partial_{j_k} \right )^k_k \delta(\vec r\,) = \frac{(-1)^k (2k+1)!!}{r^k} \left ( \hat x_{j_1} \cdots \hat x_{j_k} \right )^k_k \delta(\vec r\,).
\ee
For $k=1$ and $k=2$ these are
\be \label{delta.identity.1}
\partial_i \delta(\vec r\,) = -\frac{3 \hat x_i}{r} \delta(\vec r\,)
\ee
and
\be \label{delta.identity.2}
\left ( \partial_i \partial_j - \frac{1}{3} \delta_{i j} \partial^2 \right ) \delta(\vec r\,) = \frac{15}{r^2} \left ( \hat x_i \hat x_j - \frac{1}{3} \delta_{i j} \right ) \delta(\vec r\,) .
\ee
These can be checked using the method of Appendix~B.

%%%%%%%%%%%%%%%%%%%%%%%%%%%%%%%%%%%%%%%%%%%%%%%%%%%

\appendix

\section{Fourier transform of the Coulomb interaction}
\label{appendix1}

We first consider the Fourier transform of the Coulomb potential from coordinate space to momentum space.  Direct evaluation in spherical coordinates doing first the (trivial) $\phi$ integral and next the $\theta$ integral leads to
\be \label{CoulombFT.r.to.p}
\int d^3 r e^{-i \vec p \cdot \vec r} \, \frac{1}{4 \pi r} = \frac{1}{2} \int_0^\infty dr \, r \int_0^\pi d \theta \, \sin \theta e^{-i p r \cos \theta} = \frac{1}{2} \int_0^\infty dr \, r \left ( \frac{e^{i p r} - e^{-i p r}}{i p r} \right ) = \frac{1}{p} \int_0^\infty dr \, r \sin(pr).
\ee
(It is convenient to use the substitution $u=\cos \theta$ with limits $-1 < u < 1$ for the $\theta$ integral.)  This integral has a problem at large distances: it doesn't converge.  We repair this `infrared' divergence by the insertion of a convergence factor in the original integrand: $1/r \rightarrow e^{- \lambda r}/r$ with $\lambda$ a small positive number that will be taken to zero at the end of the calculation.  With this alteration, the transform \eqref{CoulombFT.r.to.p} becomes
\be
\int d^3 r e^{-i \vec p \cdot \vec r} \, \frac{e^{- \lambda r}}{4 \pi r} = \frac{1}{2 i p} \int_0^\infty dr \left ( e^{-(\lambda-i p)r} - e^{-(\lambda+i p)r} \right ) = \frac{1}{2 i p} \left ( \frac{1}{\lambda-i p} - \frac{1}{\lambda+i p} \right ) = \frac{1}{p^2+\lambda^2} .
\ee
On taking the $\lambda \rightarrow 0$ limit this shows that the transform of $1/(4 \pi r)$ is $1/p^2$.  The inverse transform \eqref{FTa} then takes $1/p^2$ to $1/(4 \pi r)$.

The original momentum space to coordinate space transform of Eq.~\eqref{FTa} can be performed as a contour integral.  We retain the momentum space version of the infrared convergence factor and evaluate the $\phi_p$ and $\theta_p$ integrals:
\be \label{FT.p.to.r}
\int \frac{d^3 p}{(2 \pi)^3} e^{i \vec p \cdot \vec r} \frac{1}{p^2+\lambda^2} = \frac{-i}{8 \pi^2 r} \int_0^\infty dp \left ( \frac{1}{p-i \lambda} + \frac{1}{p+i \lambda} \right ) \left ( e^{i p r} - e^{-i p r} \right ) .
\ee
We note that the integrand is even in $p$ so we can extend the integration range to $-\infty < p < \infty$ if we divide the result by 2.  We use the residue theorem to evaluation the integral, closing the $e^{i p r}$ term in the upper half plane and the $e^{-i p r}$ term in the lower half plane.  The two contributions are identical: the final result is $e^{-\lambda r}/(4 \pi r)$, as expected.  The convergence factor was convenient to move the poles off of the real $p$ axis.  Without the convergence factor the integral \eqref{FT.p.to.r} would be
\be
\int \frac{d^3 p}{(2 \pi)^3} e^{i \vec p \cdot \vec r} \frac{1}{p^2} = \frac{1}{2 \pi^2} \int_0^\infty dp \frac{\sin (pr)}{p r} = \frac{1}{4 \pi r} ,
\ee
a familiar (and tabulated) integral obtained by the one-dimensional Fourier transform of the rectangular pulse function.

%%%%%%%%%%%%%%%%%%%%%%%%%%%%%%%%%%%%%%%%%%%%%%%%%%%

\section{Verification of identities involving generalized functions}
\label{appendix2}

In this appendix we describe our means of confirming identities involving generalized functions and illustrate with proofs of several equalities from the text.  Suppose $A$ and $B$ are two generalized functions and consider the assertion $A=B$.  We adopt the `physicists' proof' approach of Frahm \cite{Frahm83} and say that $A=B$ holds if
\be \label{gen.fn.identity}
\int d^3 r \, F(\vec r\,) A = \int d^3 r \, F(\vec r\,) B
\ee
for all test functions $F(\vec r\,)$ that are smooth (infinitely differentiable) and that decrease sufficiently rapidly at large distances.  The generalized functions considered in this work have singularities at only a single point, which (by translational invariance) we have taken to be the origin.  So an identity like $A=B$ can be tested at points different from the origin by traditional means.  We let the integration region implicit in Eq.~\eqref{gen.fn.identity} include the origin, and break it up into two parts: a small sphere (of radius $R$) surrounding the origin, and an outer part.  As long as $A=B$ away from the origin the integrals over the `outer part' will be automatically identical and we needn't consider them further.  So the identity $A=B$ holds when $A=B$ away from the origin and
\be \label{gen.fn.identity.2}
\int_{B(R)} d^3 r \, F(\vec r\,) A = \int_{B(R)} d^3 r \, F(\vec r\,) B
\ee
in the limit $R \rightarrow 0$ where the integration region is the ball $B(R)$ of radius $R$ surrounding the origin.  The integral over $B(R)$ can be written as
\be
\int_{B(R)} d^3 r = \int_0^R dr \, r^2 \int d \Omega ,
\ee
where $d \Omega$ is the element of solid angle, and by convention we always perform the angular integrals first.  For some of the identities to be considered, one of the generalized functions is a gradient.  Suppose then that $A=\partial_k \Lambda$.  We evaluate the integral over $B(R)$ according to
\bearray \label{form.for.gradient}
\int_{B(R)} d^3 r \, F(\vec r\,) \partial_k \Lambda &=& \int_{B(R)} d^3 r \, \partial_k  \left ( F(\vec r\,) \Lambda \right ) - \int_{B(R)} d^3 r \left ( \partial_k F(\vec r\,) \right ) \Lambda \cr
&=& \oint_{S(R)} d \Omega \, R^2 \hat x_k F(\vec r\,) \Lambda - \int_{B(R)} d^3 r \left ( \partial_k F(\vec r\,) \right ) \Lambda ,
\eearray
where the integral in the first term is over the sphere $S(R)$ of radius $R$ that is the surface of $B(R)$.  The surface element of $S(R)$ is $d \vec S = dS \hat r = d \Omega \, R^2 \hat r$.   The integral identity used in transforming the first term in Eq.~\eqref{form.for.gradient} is a generalization of Gauss's Theorem. \cite{Arfken01d}  The test functions $F(\vec r\,)$ are infinitely differentiable and so can be expanded in Taylor series around the point $\vec r=0$:
\be
F(\vec r\,) = F_0 + x_a F_a + \frac{1}{2} x_a x_b F_{a b} + \cdots
\ee
where $F_0=F(0)$, $F_a=\partial_a F(0)$, $F_{a b} = \partial_a \partial_b F(0)$, etc.  It follows that $\partial_k F(\vec r\,) = F_k + x_a F_{a k} + \frac{1}{2} x_a x_b F_{a b k} + \cdots$.

In Sec.~\ref{sec2} it was asserted that the three-dimensional delta function can be written as
\be \label{delta.equality}
\delta(\vec r\,) = \frac{\delta(r)}{4 \pi r^2} .
\ee
The proof of this statement is now straightforward.  The left hand side (LHS) of Eq.~\eqref{gen.fn.identity} is simply $F(0)$ by use of the sifting property of the three-dimensional delta function.  The right hand side (RHS) of Eq.~\eqref{gen.fn.identity} is
\be
{\rm RHS} = \frac{1}{4 \pi} \int_0^R dr \, \delta(r) \int d \Omega \, F(r,\theta,\phi) = F(0) ,
\ee
since the one-dimensional delta function has unit area in the region $0 < r \le R$ for any positive $R$ and $F(0,\theta,\phi)=F(0)$ for any $\theta$, $\phi$.  The equality \eqref{delta.equality} is thus confirmed.

As a second example, we consider the proof of Eq.~\eqref{onebyrsq.2}.  In this case the left hand side of Eq.~\eqref{onebyrsq.2} is a gradient $\partial_k \Lambda_k$ where 
\be
\Lambda_k = \left ( \delta_{i k} \partial_j - \frac{1}{3} \delta_{i j} \partial_k \right ) \frac{1}{r^2} = -\frac{2}{r^3} \left ( \delta_{i k} \hat x_j - \frac{1}{3} \delta_{i j} \hat x_k \right ) .
\ee
The first term on the LHS of Eq.~\eqref{gen.fn.identity} (the first term of Eq.~\eqref{form.for.gradient}) is
\be
{\rm LHS}_1 = \oint d\Omega \, R^2 \hat x_k F(\vec r\,) \left ( \frac{-2}{R^3} \right ) \left ( \delta_{i k} \hat x_j - \frac{1}{3} \delta_{i j} \hat x_k \right ) = \frac{-2}{R} \oint d \Omega \, \left ( F_0 + x_a F_a + \frac{1}{2} x_a x_b F_{a b} + \cdots \right ) \left ( \hat x_i \hat x_j \right )^2_2 .
\ee
Now the $\left ( \hat x_i \hat x_j \right )^2_2$ term has angular momentum $\ell=2$ and will vanish when multiplied by $F_0$ (which has $\ell=0$) or $x_a$ (which has $\ell=1$) and averaged over angles.  The $x_a x_b F_{a b}$ term is the first in the Taylor expansion to survive the angular average, but it contributes at order $R$ and so disappears in the $R \rightarrow 0$ limit.  Higher order terms in the expansion of $F(\vec r\,)$ are even higher order in $R$ and so do not contribute either.  The second term on the LHS of Eq.~\eqref{gen.fn.identity} (the second term of Eq.~\eqref{form.for.gradient}) is
\be
{\rm LHS}_2 = -\int_0^R dr r^2 \int d\Omega \left ( F_k + r \hat x_a F_{a k} + \cdots \right ) \left ( \frac{-2}{r^3} \right ) \left ( \delta_{i k} \hat x_j - \frac{1}{3} \delta_{i j} \hat x_k \right ) .
\ee
The angular average eliminates the product of $F_i$ (which has $\ell=0$) with $\hat x_j$ or $\hat x_k$ (which both have $\ell=1$).  The next term in the series for $F(\vec r\,)$ contributes at order $R$ and so vanishes as do all higher terms in the expansion.  The RHS of Eq.~\eqref{gen.fn.identity} is
\be
{\rm RHS} = \int_0^R dr \, r^2 \int d\Omega \, F(\vec r\,) \frac{8}{r^4} \left ( \hat x_i \hat x_j \right )^2_2 .
\ee
The first term in the expansion for $F(\vec r\,)$ that survives the angular averaging is $\frac{r^2}{2} \hat x_a \hat x_b F_{a b}$, which leads to a negligible contribution of order $R$.  Identity \eqref{onebyrsq.2} holds for $r>0$ by explicit calculation, and it holds near $r=0$ because ${\rm LHS} = {\rm RHS}$ (both being 0), so Eq.~\eqref{onebyrsq.2} is an identity among generalized functions.

As a final example, we consider the delta function identity given by Eq.~\eqref{delta.identity.2}.  The $\Lambda_k$ quantity for this case,
\be
\Lambda_k = \left ( \delta_{i k} \partial_j - \frac{1}{3} \delta_{i j} \partial_k \right ) \delta(\vec r\,) = -\frac{3}{r} \left ( \delta_{i k} \hat x_j - \frac{1}{3} \delta_{i j} \hat x_k \right ) \delta(\vec r\,) ,
\ee
itself contains a delta function and so vanishes unless $r=0$.  It follows that the surface term, ${\rm LHS}_1$, is zero.  The second LHS term is
\be
{\rm LHS}_2 = -\int d^3 r \left ( \partial_k F(\vec r\,) \right ) \Lambda_k = -\int d^3 r \, \left ( F_k + x_a F_{a k} + \cdots \right ) \left ( \frac{-3}{r} \right ) \left ( \delta_{i k} \hat x_j - \frac{1}{3} \delta_{i j} \hat x_k \right ) \delta(\vec r\,) = F_{i j} - \frac{1}{3} \delta_{i j} F_{k k} .
\ee
A similar calculation shows that the RHS is also $F_{i j}-\frac{1}{3} \delta_{i j} F_{k k}$, completing the proof.

%%%%%%%%%%%%%%%%%%%%%%%%%%%%%%%%%%%%%%%%%%%%%%%%%%%

%%%%%%%%%%%%%%%%%%%%%%%%%%%%%%%%%%%%%%%%%%%%%%%%%%%


\begin{thebibliography}{10}
\bibitem{note1} The angular part $d \Omega_p$ of the momentum space volume element can be identified from writing $d^3 p = dp p^2 d \Omega_p$ where $d \Omega_p$ is the element of solid angle in momentum space: $d \Omega_p = d \theta_p \sin \theta_p d \phi_p$ with $\hat p = \vec p/p = ( \sin \theta_p \cos \phi_p, \sin \theta_p \sin \phi_p, \cos \theta_p)$.
\bibitem{note2} When $\ell$ is not an odd integer we use $(2\ell+1)!! \equiv 2^{\ell+1} \Gamma(\ell+3/2)/\Gamma(1/2)$.
\bibitem{note3} We do not adopt the suggestion of Blinder (Ref.~\onlinecite{Blinder03}) and Hu (Ref.~\onlinecite{Hu04}) that the radial delta function should be defined with area $1/2$ instead of area $1$.
\bibitem{note4} We do not accept the requirement proposed by Menon (Ref.~\onlinecite{Menon99}) that $\lim_{\lambda \rightarrow 0}  \left ( \lim_{r \rightarrow 0} R_\ell(\lambda; r) \right )$ is necessarily infinite.  See also the discussion by Hnizdo (Ref.~\onlinecite{Hnizdo00}).
\end{thebibliography}
\end{document}